\documentclass[conference]{IEEEtran}

\IEEEoverridecommandlockouts
\usepackage{subfigure} 
\usepackage{array}
\usepackage{algorithm}
\usepackage{algpseudocode}
\usepackage{cite}
\usepackage{amsmath,amssymb,amsfonts}
\usepackage{graphicx}
\usepackage{textcomp}
\usepackage{xcolor}
\usepackage{float}
\usepackage{tabularx}
\usepackage{multirow}
\usepackage{adjustbox}
\usepackage{svg}
\usepackage{balance}

\usepackage{hyperref} 

\def\BibTeX{{\rm B\kern-.05em{\sc i\kern-.025em b}\kern-.08em
    T\kern-.1667em\lower.7ex\hbox{E}\kern-.125emX}}

\begin{document}
\pagestyle{empty}

\title{{\fontsize{23pt}{12pt}\selectfont Modelling the 5G Energy Consumption Using Real-world Data: Energy Fingerprint is All You Need}}

    
\author{Tingwei~Chen, 
        Yantao~Wang, 
        Hanzhi~Chen, 
        Zijian~Zhao, 
        Xinhao~Li, 
        Nicola~Piovesan,~\IEEEmembership{Senior Member,~IEEE,}\\
        Guangxu~Zhu,~\IEEEmembership{Senior Member,~IEEE,}
        and~Qingjiang~Shi,~\IEEEmembership{Senior Member,~IEEE}%

\thanks{Tingwei Chen, Yantao Wang, Hanzhi Chen, Zijian Zhao, Xinhao Li, Guangxu Zhu, and Qingjiang Shi are with the Shenzhen Research Institute of Big Data, Shenzhen, China. (email: tingweichen@link.cuhk.edu.cn; wangyantao74748999@gmail.com; hanzhi001@e.ntu.edu.sg; zhaozj28@mail2.sysu.edu.cn; xinhaoli@link.cuhk.edu.cn; gxzhu@sribd.cn; shiqj@sribd.cn).}%
\thanks{Hanzhi Chen is also with the School of Electrical and Electronic Engineering, Nanyang Technological University, Singapore, Singapore.}%
\thanks{Zijian Zhao is also with the School of Computer Science and Engineering, Sun Yat-sen University, Guangzhou, China.}%
\thanks{Xinhao Li is also with the School of Science and Engineering, The Chinese University of Hong Kong (Shenzhen), Shenzhen, China.}%
\thanks{Nicola Piovesan is with Huawei Technologies, Boulogne-Billancourt, France (email: nicola.piovesan@huawei.com).}%
\thanks{Qingjiang Shi is also with the School of Software Engineering, Tongji University, Shanghai, China.}%
\thanks{Corresponding author: Guangxu Zhu (email: gxzhu@sribd.cn).}}%

\maketitle
\thispagestyle{empty}

\begin{abstract}

The introduction of 5G technology has revolutionized communications, enabling unprecedented capacity, connectivity, and ultra-fast, reliable communications. However, this leap has led to a substantial increase in energy consumption, presenting a critical challenge for network sustainability. Accurate energy consumption modeling is essential for developing energy-efficient strategies, enabling operators to optimize resource utilization while maintaining network performance. To address this, we propose a novel deep learning model for 5G base station energy consumption estimation based on a real-world dataset. Unlike existing methods, our approach integrates the Base Station Identifier (BSID) as an input feature through an embedding layer, capturing unique energy patterns across different base stations. We further introduce a masked training method and an attention mechanism to enhance generalization and accuracy. Experimental results show significant improvements, reducing Mean Absolute Percentage Error (MAPE) from 12.75\% to 4.98\%, achieving over 60\% performance gain compared to existing models. The source code for our model is available at \href{https://github.com/RS2002/ARL}{GitHub Repository}.

\end{abstract}
\section{Introduction}
The fifth-generation (5G) network has revolutionized communication technologies, enabling transformative applications such as virtual reality, vehicular networks, and the Internet of Things (IoT). However, this technological advancement has also led to a significant increase in energy consumption. Compared to the fourth-generation (4G) network, 5G networks consume approximately three times more energy \cite{green5g2020}. For mobile network operators, energy costs now represent around 22.5\% of total operational expenses \cite{gsma5g2020}, and this proportion continues to grow. A major contributor to this energy consumption is the radio access network (RAN), with base stations (BSs) alone accounting for over 70\% of the total network energy usage \cite{9678321}.

Although base stations (BSs) are inherently energy-intensive, their energy consumption can be optimized by dynamically disabling certain hardware components based on traffic load. Accurate energy consumption modeling is crucial for implementing such energy-saving strategies. Previous studies have extensively investigated the impact of BS architectures and configurations on power consumption. For instance,~\cite{auer2011much} proposed an analytical model demonstrating a linear relationship between BS output and power consumption. This model was later extended in~\cite{debaillie2015flexible} to incorporate massive multiple-input multiple-output (mMIMO) architectures and energy-efficient techniques. Further refinements were made in~\cite{tombaz2015energy}, which accounted for the linear increase in power consumption with additional mMIMO transceivers. A comprehensive framework was also introduced in~\cite{bjornson2015optimal}, considering factors such as mMIMO architecture, downlink and uplink phases, and the number of multiplexed users. Additionally, the 3GPP technical report TR 38.864 \cite{3GPP_TR_38_864} provides a standardized energy consumption model for 5G networks, focusing on energy-saving techniques such as dynamic resource allocation and power amplifier efficiency optimization. However, these models fail to fully capture the complexity of real-world 5G radio units, particularly the heterogeneity of hardware and environmental factors.

Most recently, SRCON (Simulated Reality of Communication Networks) has emerged as a revolutionary paradigm in network simulation. Detailed in ~\cite{luo2023srcon,lopez2023data}, SRCON leverages an innovative mix of white-box and black-box models to mimic the stochastic behaviors observed in real-world 4G/5G mobile networks, offering a robust framework for accurate simulation. Within this framework,~\cite{piovesan2022machine,piovesan2023power} develops a machine learning model, trained on extensive network measurement data, which significantly improves the accuracy of energy consumption estimations compared to traditional mathematical modeling.

However, these approaches typically overlook the domain shifts caused by variations in hardware configurations and physical environments. For instance, different base stations may have completely distinct equipments and physical environments, yet such differences are not adequately represented in the input features of current models. This domain discrepancy inevitably leads to prediction biases. The observed phenomenon - where data from the same base station shows consistent patterns while significant variations exist across different stations.

To address these limitations, we propose a data-driven approach based on a real-world 5G radio unit dataset from the ITU 5G Base Station Energy Consumption Modelling Challenge~\cite{ZindiAIML5G2023}. The dataset exhibits a characteristic domain shift problem, where identical input features correspond to varying energy consumption patterns across different base stations. Unlike existing methods, our approach incorporates the Base Station Identifier (BSID) as a key feature to capture the domain-specific characteristics. We utilize an embedding layer to represent the BSID, which provides a more compact representation compared to sparse one-hot encoding. To further improve robustness, we introduce a masked training technique to handle unseen BSIDs in test data. Additionally, an attention mechanism is employed to automatically select the most relevant features for energy consumption estimation. Experimental results demonstrate that our approach reduces the Mean Absolute Percentage Error (MAPE) from 12.75\% to 4.98\%, achieving a performance improvement of over 60\% compared to state-of-the-art methods~\cite{piovesan2023power}.


\section{5G Radio Unit Dataset}






In this section, we introduce the 5G Radio Unit Dataset used in our study. The dataset consists of 102,705 hourly measurements collected over a period of 8 days. Each sample includes a variety of features, as detailed in Table~\ref{table:engineering_parameters}, which are classified into three main categories along with energy consumption:

\begin{itemize}

\item \textbf{Base Station features:} 
These include hardware attributes and configuration details of the BS, which are key in predicting energy consumption. Features in this category are the number of antennas (Antennas), the transmission mode (Mode), the BSID, and the type of radio unit (RUType). BSs with similar hardware attributes and configurations often show similar energy consumption patterns.

\item \textbf{Cell-Level features:} 
These include hardware attributes and configuration details of each of the cells operated by the BS.
Features in this category are Load, activation levels of six different Energy Saving Modes (ESMode1-6), Frequency, Bandwidth, and maximum transmit power (TXpower). These dynamic features are crucial for understanding the hourly energy consumption of the BSs. Additionally, a BS can support up to four cells, with one serving as the primary and the others almost always remaining inactive.



\item \textbf{Time features:} The dataset includes day and hour fields, indicating when the measurement samples were collected. 

\item \textbf{Energy Consumption:} The dataset includes hourly measurements of energy consumption.

\end{itemize}

The dataset is split into training and testing sets to test our model's generalization capabilities, with certain BSs only present in the testing set. The training set consists of 92,629 samples. The testing set includes 10,076 samples, of which 3,067 are from BSs not encountered during training (\textit{cross-domain}), and 7,009 are from BSs that appear in the training set (\textit{in-domain}).



\section{Problem Statement}
\begin{figure}[!t]
\centering
\includegraphics[trim={0 0 1cm 1cm},clip,width=0.4\textwidth]{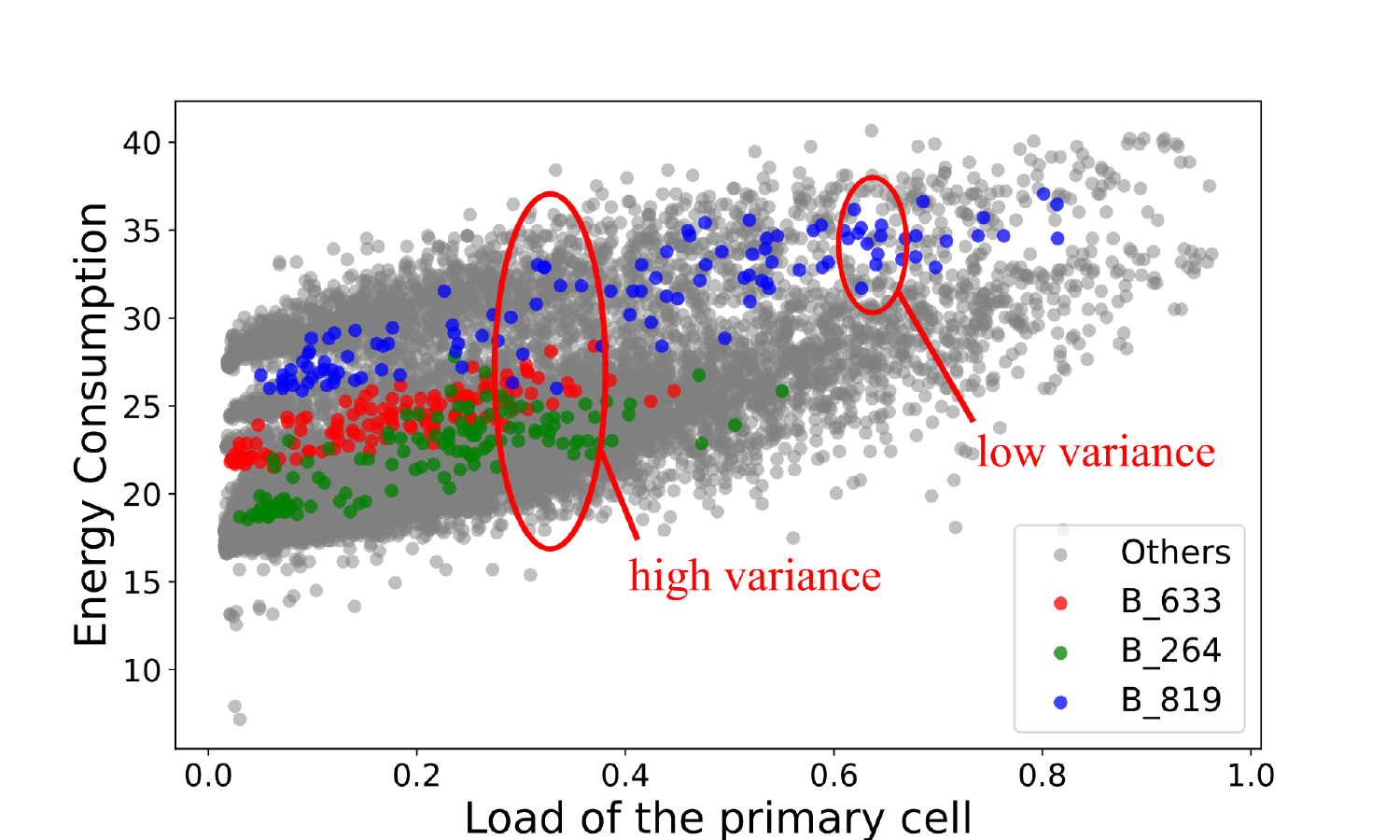}
\caption{Illustration of the “one-to-many" issue. Different colors represent sample points from different BS while sharing the same input features.}
\label{problem}
\end{figure}

By analysis of the collected data, there exists a “one-to-many" issue in the datasets, meaning that different energy consumption values are reported for identical input feature values. This issue arises from two primary origins: i) the absence of certain unknown features, leading to a loss of information; ii) errors that occur during the measurement of energy consumption. Fig. 1 is a scatter plot showing the energy consumption values of selected data samples plotted against the load of the primary cell, which is identified as one of the most important features in \cite{piovesan2023power}. Three key observations can be made from Fig.~\ref{problem} as listed below:
\begin{itemize}
    \item Energy consumption exhibits a positive correlation with the load of the primary cell, where higher loads tend to correspond to higher energy consumption.
    \item Without distinguishing BSID, as depicted by the gray area in the figure, a common load value corresponds to a high variance in energy consumption.
    \item When distinguishing by BSID, as depicted by the red, green, and blue areas, a common load value corresponds to much lower variance in energy consumption.
\end{itemize}

Combining these observations, we conclude that distinguishing BSID enhances our ability to identify patterns in energy consumption variations across different BSs. This gain comes from the fact that incorporating BSID as a model input feature enables us to unveil the subtle hardware differences across various BSs, effectively capturing the unique BS energy fingerprint for more accurate predictions. 

To mathematically model the relationship between input features and energy consumption, we define a mapping function \( f \) where \( y \in \mathbb{R} \) denotes the energy consumption, and \( \boldsymbol{x} \in \mathbb{R}^n \) denotes the input features, formed by concatenating various features and BSID, as shown in Table~\ref{table:engineering_parameters}. Since deriving \( f(\boldsymbol{x}) \) analytically is challenging, we propose using a neural network \( f(\boldsymbol{x}; \theta) \) with trainable parameters \( \theta \):

\begin{equation}
    y = f(\boldsymbol{x}; \theta).
\end{equation}
By including BSID in $\boldsymbol{x}$, the model captures unique base station characteristics, reducing uncertainty from unknown factors.

In practical deployment scenarios, the proposed method allows for the use of alternative features during local training, rather than being limited to the features used in the dataset. The requirements for the amount of training samples can be referenced in \cite{piovesan2023power}.

\section{Deep Learning Model Architecture and Training}
In this section, we present our methodology, which includes input processing techniques, the architecture of our deep learning model, and a novel masked training approach designed to improve generalization across different BSs. This methodology addresses the one-to-many issues outlined in Section III.
\subsection{Input Feature Processing}
Our input feature processing employs distinct encoding methods to handle the diversity of input features effectively. Categorical inputs (excluding BSID) are processed using One-Hot encoding to preserve categorical distinctions. Numerical inputs are used directly to leverage their magnitude variations, except for features like Antennas, Bandwidth, and Frequency, which have limited value ranges and are One-Hot encoded to capture subtle variations. For BSID, which has a large number of classes, we use embedding techniques to transform it into a compact numerical vector, balancing dimensionality and information retention. This tailored approach has significantly improved model performance, as discussed in Section IV. The encoding methods for all input features are summarized in Table~\ref{table:engineering_parameters}.


\begin{table}[!t]

\centering
\caption{Input Features.}
\begin{tabular}{cccc}
\hline
\textbf{Feature Class}                & \textbf{Feature}                                    & \textbf{Type}    &\textbf{Feature Encoding}   \\ \hline
BS-level  & RUType           & Categorical &One-Hot\\ 
                      & Mode                     & Categorical &One-Hot\\ 
                      & Antennas  & Numerical &One-Hot\\ 
                      & BSID                                         & Categorical &Embedding
                      \\ \hline
Cell-level   &Load & Numerical&Direct Input\\
&ESMode1-6& Numerical&Direct Input\\& TXpower    & Numerical  & Direct Input\\
& Frequency               & Numerical &One-Hot\\ 
                      & Bandwidth               & Numerical&One-Hot \\ 
                       \hline
Time                      & Day                                             & Categorical &One-Hot\\ 
                      & Hour                                            & Categorical &One-Hot\\ \hline
                      
\end{tabular}
\label{table:engineering_parameters}
\end{table}

\subsection{Deep Learning Model Design}
\begin{figure*}[ht]
\centering
\includegraphics[trim={0cm 0cm 0cm 1cm},clip,width=1\textwidth]{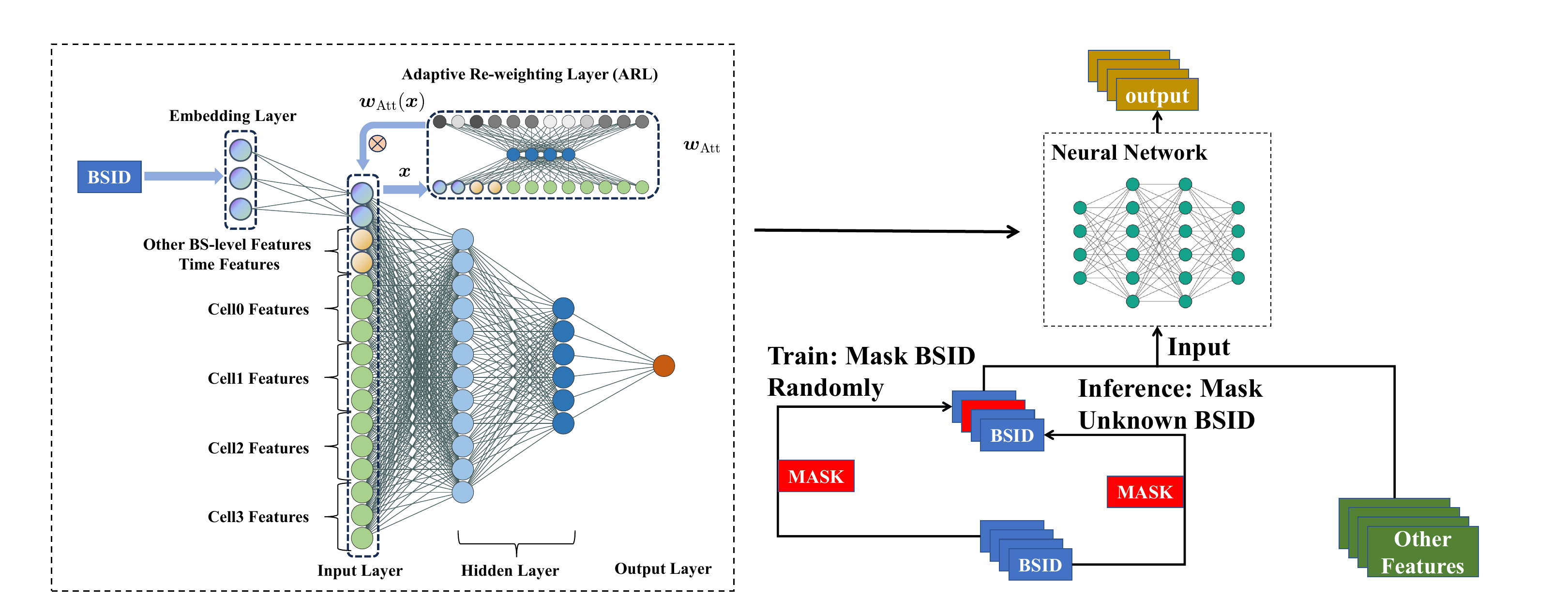}

\caption{Model Architecture: Our model is a two-layer Feed Forward Network. In the training phase, random masking is applied to randomly reassign BSIDs to `Unknown BS'.}
\label{overview}
\end{figure*}

Fig.~\ref{overview}(left) illustrates the architecture of our proposed framework.
At its core, the model employs a multilayer perceptron (MLP) structure, and includes specialized modules for embedding and adaptive weight adjustment. 
The model consists of two hidden layers: the first layer contains 128 neurons, while the second layer comprises 64 neurons. The final layer directly outputs the predictions.

\subsubsection*{\textbf{BSID Embedding}}
The efficient processing of BSID is a critical aspect, especially considering the large number of BSs included in our dataset.
The use of traditional One-Hot encoding for BSIDs produces sparse vectors, which can become inefficient.

To tackle this challenge, we have integrated an embedding technique into our model. This approach transforms each categorical BSID into a dense, continuous vector. Specifically, each BSID is represented as an index $\textrm{BSID}$ in a lookup table $\boldsymbol{W}_{\textrm{emb}}$. Initially, each embedding is randomly assigned a compact 64-dimensional numerical vector, $ \boldsymbol{x}_{\textrm{BSID}}$. 
During training, automatic differentiation updates embedding values directly, refining them implicitly without converting indices to One-Hot vectors, thereby optimizing embeddings efficiently. This transformation is mathematically expressed as:
\begin{equation}
    \boldsymbol{x}_{\textrm{BSID}} =\boldsymbol{W}_{\textrm{emb}}[\textrm{BSID}].
\end{equation}


This embedding process, which involves learning the values of $\boldsymbol{W}_{\textrm{emb}}$ during training, effectively reduces both sparsity and dimensionality, thus simplifying the model and mitigating its computational demand. Additionally, the embedding allows the model to capture more complex relationships within the categorical BSID.





\subsubsection*{\textbf{Attention mechanism}}

To enhance feature selection and identify key elements within the feature vector, we propose an optimized network structure with a novel attention mechanism, the adaptive re-weighting layer (ARL). ARL improves predictions by assigning different weights to each feature element. The module first compresses the input, $x$ via a linear layer, then expands it back to the original dimensionality. A rectified linear unit (ReLU) function is applied in the intermediate stage, followed by a sigmoid activation, producing a vector of values between 0 and 1 to re-weight the input features. This process is described by the following equation:
\begin{equation}
\boldsymbol{w}_{\textrm{Att}}(\boldsymbol{x})=\text{Sigmoid}(\text{Linear}(\text{ReLU}(\text{Linear}(\boldsymbol{x})))).
\end{equation}


The attention vector, $\boldsymbol{w}_{\textrm{Att}}(x)$, is then applied to the original input through element-wise multiplication to obtain the re-weighted input of the network, $\boldsymbol{x}{'}$, as shown below:
\begin{equation}
    \boldsymbol{x}{'}=\boldsymbol{w}_{\textrm{Att}}(\boldsymbol{x})\cdot \boldsymbol{x}.
\end{equation}
This process ensures that the network assigns higher significance to the most relevant features, thereby enhancing its precision and effectiveness.



\subsection{Training with Random BSID Masking}
To enhance our model's generalization across unknown BSs, we have integrated a distinct BSID category for such BSs. As illustrated in Fig.~\ref{overview}(right), during the training phase, 30\% of the samples in each epoch are randomly assigned this `unknown BS' identifier. This technique is designed to enable the model to handle unknown BSIDs during the inference phase, enhancing its ability to generalize and avoid overfitting to BSs in the test set. 

Our model's training objective is to minimize the MAPE, which is defined as follows.
\begin{equation}\label{metric}
\textrm{MAPE} = \frac{\sum_{i=1}^n \left| {y_i} - \widehat{y}_i \right|}{\sum_{i=1}^n \left| {y_i} \right|},
\end{equation}
where $y_i$ represents the i-th measured energy consumption sample, and $\widehat{y}_i$ denotes the model's corresponding estimation. 

The model was trained for 1000 epochs with a batch size of 512. After each epoch, the model was evaluated on the test set, and the epoch with the best test performance was selected as the final model. This approach ensures that the model's performance is reported under its optimal conditions.

In practical eployment scenarios, the masked training strategy can be employed for datasets with different input features. During training, the model alternates between using and omitting specific feature values across epochs, enabling adaptation to unseen values while maintaining performance on observed values. This enhances generalization to new or incomplete data during inference.

\section{Experiment and Analysis}

In this section, we present the performance achieved by our model and a comprehensive analysis of experiments conducted to benchmark the performance of our model against a range of alternative methods.


\subsection{Overall performance}
Our model was trained for over 1000 epochs. The trained model was tested on 10,076 samples from the testing set and achieved remarkable performance, with a MAPE value of 4.9811\%.
\subsection{BSID Generalization Capabilities}

In this section, we discuss strategies to enhance our model's generalization across different BSs by varying the processing of BSID. We compare two approaches: using One-Hot encoded BSID versus embedded BSID alongside other input features.

As shown in Table~\ref{TABLE_COMBINED}, using One-Hot encoded BSID resulted in significant predictive errors for unknown BSs. Without masked training, the model performed worse than omitting BSID entirely. However, masked training improved generalization for unknown BSs while maintaining accuracy for known BSs. Further, combining BSID embedding with masked training significantly boosted accuracy for unknown BSs, outperforming One-Hot encoding. For comparison, we also tested the method proposed by~\cite{piovesan2022machine}, which omits BSID and uses a different encoding approach, but it performed poorly on this dataset.

\begin{figure}[ht]
\centering
\includegraphics[trim={2.2cm 1.5cm 2cm 2.44cm},clip,width=0.35\textwidth]{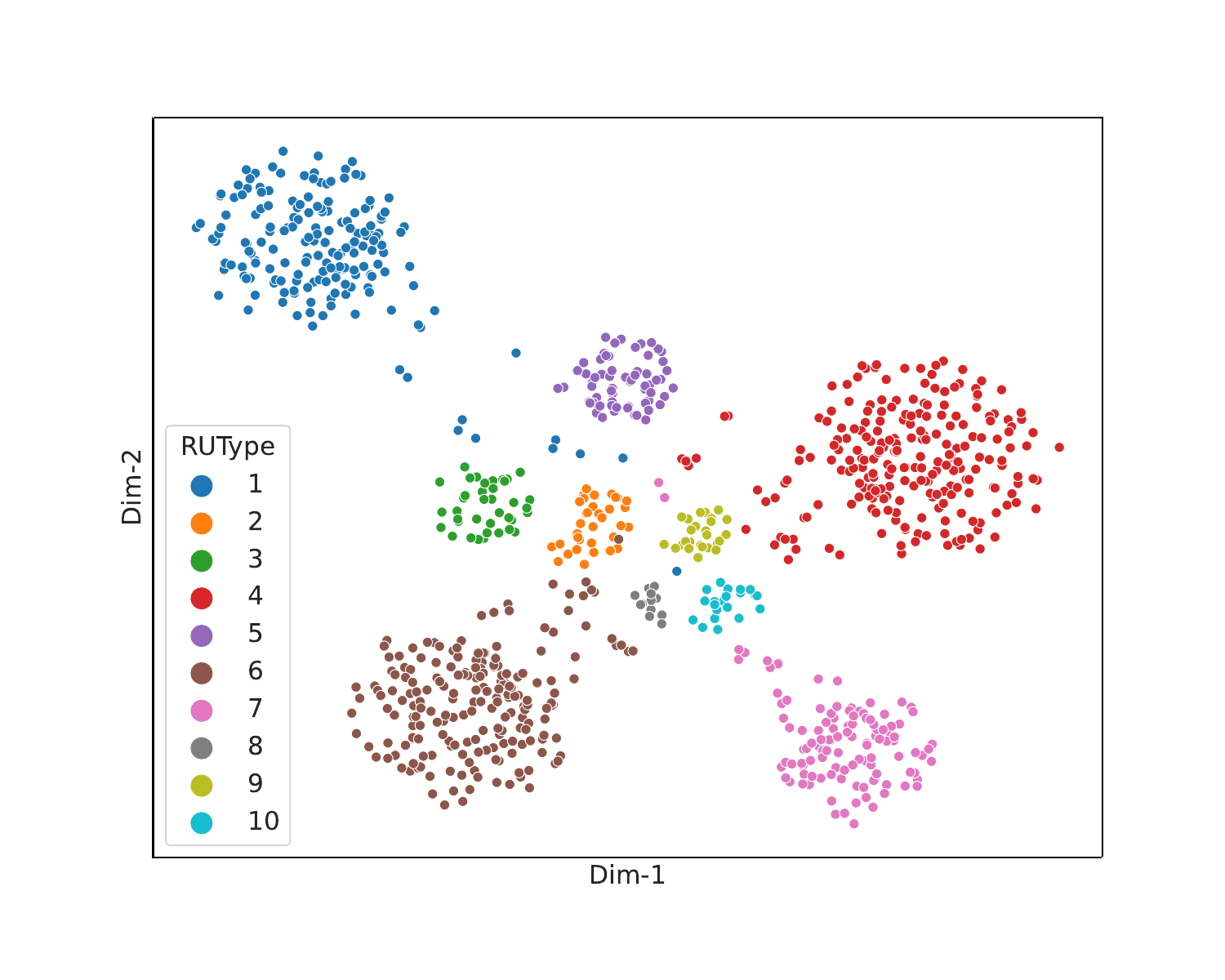}
\caption{2-Dimension visualization of BSID embeddings via UMAP. Different colors represent different RUTypes.}
\label{BS_embedding}
\end{figure}

To analyze these results, we used UMAP for dimensionality reduction, compressing BS embeddings into 2D for visualization. Fig.~\ref{BS_embedding} shows that BSs of the same RUType cluster together in the embedding space, indicating that the embedding effectively captures structural similarities.

\begin{figure*}
\centering 
\subfigure[Without BSID (left) and with BSID embedding (right)]{\includegraphics[trim={0cm 0.9cm 3cm 0cm},clip,width=0.64\textwidth]{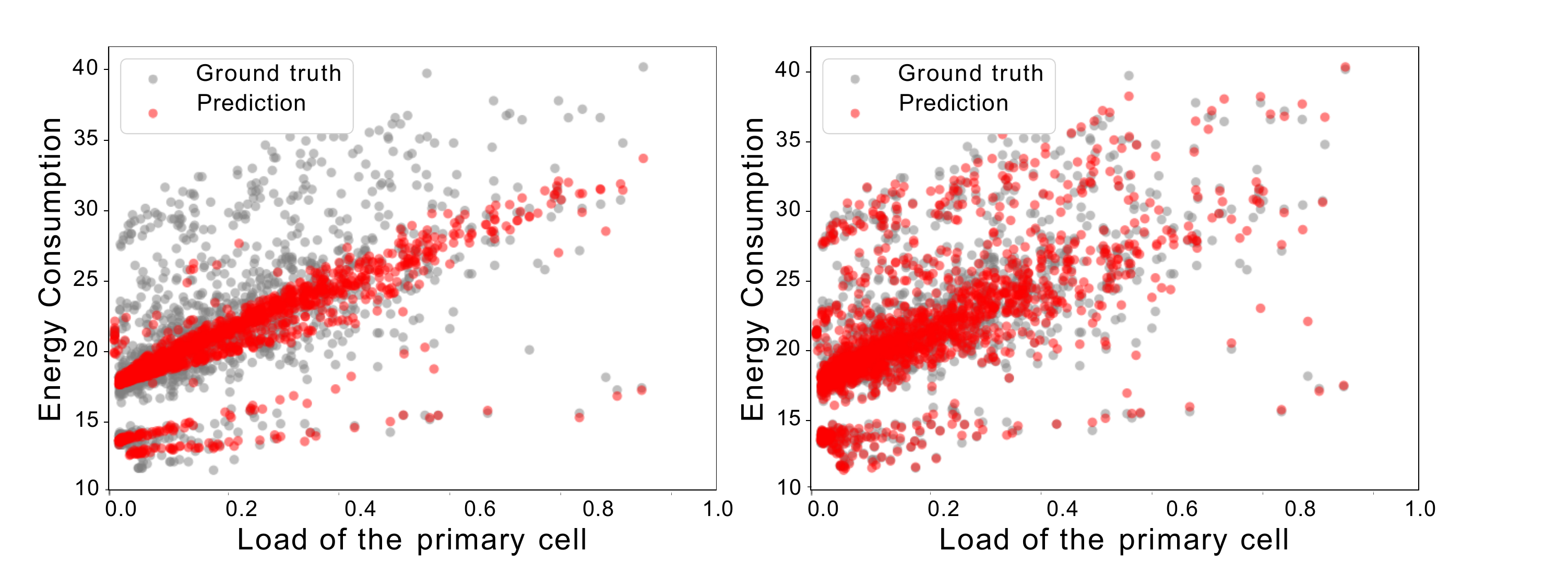}}
\subfigure[Different encoding methods for ABF]{\includegraphics[trim={0cm 0cm 2cm 1.8cm},clip,width=0.30\textwidth]{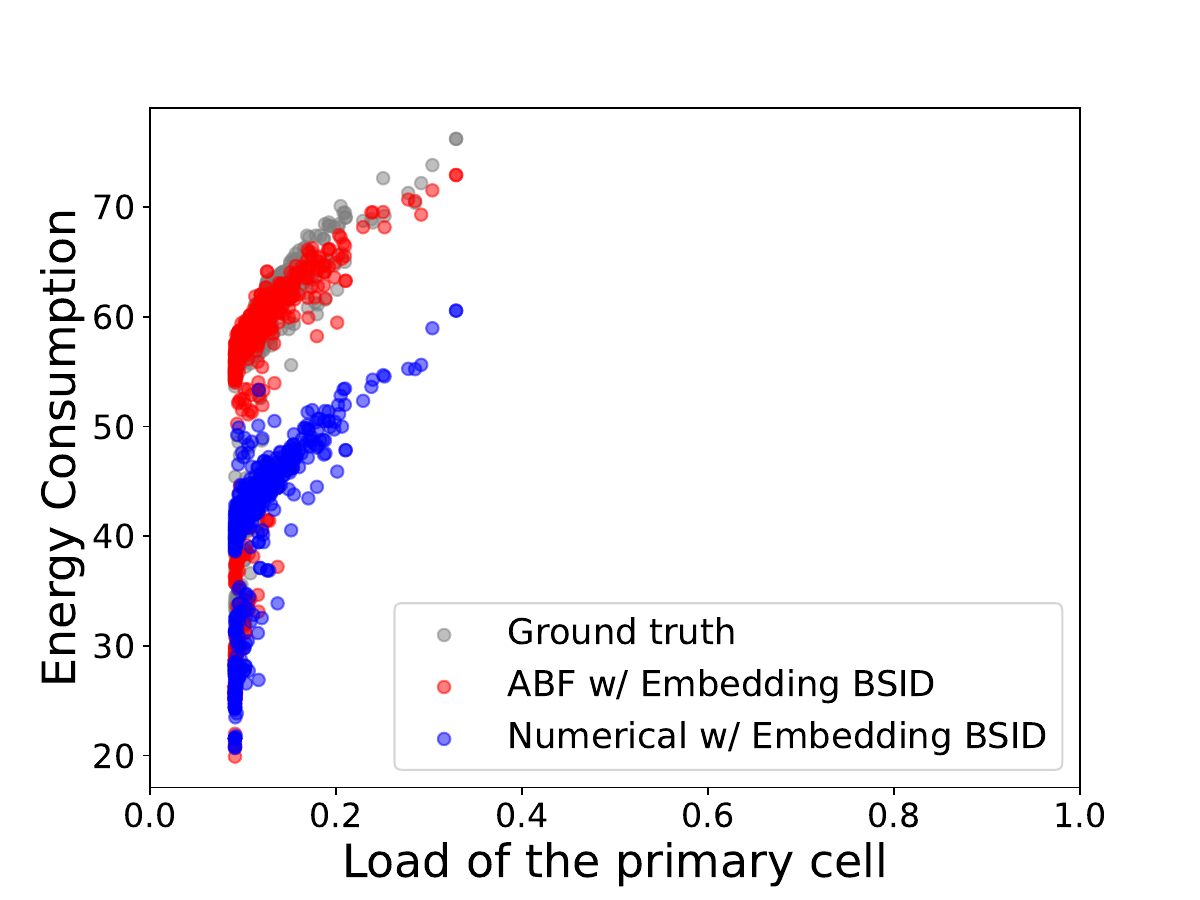}}
\caption{Comparative visualization of predictions for a known RUType dataset: (a) Model without BSID, model with embedding BSID, and (b) One-Hot vs. Numerical encoding for ABF on an unknown RUType dataset.}
\label{fig:RUType6_comparative}
\end{figure*}

Fig.~\ref{fig:RUType6_comparative}(a) compares energy consumption predictions for a known RUType with and without BSID embedding. When BSID is omitted, the model converges to a generalized pattern, predicting energy consumption based on the average behavior of all BSs for that RUType. In contrast, incorporating BSID embedding allows the model to capture the unique characteristics of individual BSs, significantly improving prediction accuracy.

\subsection{Performance Comparison of One-Hot Encoding Combinations}

In this section, we evaluate our model's performance by exploring different methods for incorporating numerical features such as frequency, bandwidth, and the number of antennas. Although these features are inherently numerical, their values in the dataset are sparse (i.e., they have limited and non-dense distributions). This sparsity makes it challenging to establish a clear linear relationship between these features and energy consumption across all BSs, potentially impacting model accuracy.

Given the sparse nature of these features, we investigated whether treating them as categorical variables, rather than numerical, could improve performance. Specifically, we compared their treatment as numerical features versus categorical features using One-Hot encoding. The goal was to identify the configuration that achieves the lowest MAPE. Table \ref{TABLE_COMBINED} summarizes the results, where “A”, “B”, and “F” represent Antennas, Bandwidth, and Frequency, respectively. Configurations such as “ABF”, “BF”, “AF”, “AB”, “A”, “B”, and “F” indicate categorical treatment with One-Hot encoding. The “ABF” configuration emerged as the most effective, achieving the lowest MAPE for both known and unknown BSs. This outcome underscores the advantage of using One-Hot encoding for sparse numerical features.

Furthermore, Fig.~\ref{fig:RUType6_comparative}(b) visualizes energy consumption predictions for a new RUType, comparing numerical and One-Hot encoding. The results demonstrate that categorical encoding aligns more closely with groundtruth data, particularly for sparse features like ABF. This confirms that the sparsity of these numerical features in the dataset is a key factor driving the superior performance of One-Hot encoding in this context.

\subsection{Evaluating the Effectiveness of Attention Mechanism}
In this section, we assess the efficacy of our proposed ARL through an ablation study. We compare two network configurations: one with ARL and one without. The results, shown in Table~\ref{TABLE_COMBINED}, reveal that ARL significantly improves MAPE across all BSs, with only a slight increase in model parameters. Notably, ARL reduces MAPE for unknown BSs from 8.5095\% to 6.6863\%, highlighting its ability to enhance generalization. To evaluate that performance gains are due to increased parameters, we compare ARL to a deeper MLP model (256-128-64). Results show that simply increasing model depth does not significantly improve performance, confirming that ARL's gains stem from its unique design rather than model scaling.





\begin{table}[!t]
\caption{Comparison of BSID encoding methods, One-Hot encoding combinations, and models with attention mechanisms on MAPE (\%). Best performances are highlighted in \textbf{bold}.}
\centering
\begin{adjustbox}{width=0.5\textwidth}
\begin{tabular}{c|c|c|c|c}
\hline
 \multicolumn{5}{c}{\textbf{BSID Encoding Methods}} \\
\hline
 Methods (Dim) & Cross-domain & In-domain & Average \\
\hline
Embedding BSID w/ RM(204) (Ours)&\textbf {6.6863}&4.0& \textbf {4.9811}\\
Embedding BSID w/o RM(204)&21.8093& 3.9984&11.4294\\
One-Hot BSID w/ RM(1064)&9.44&\textbf {3.9452}&6.1313\\
One-Hot BSID w/o RM(1064)&20.9739&4.0732&11.2645\\
No BSID(140)&8.1589&9.2191&8.7538\\
Baseline\cite{piovesan2022machine}(55)&13.9083&11.3856&12.7587\\
\hline
\multicolumn{5}{c}{\textbf{One-Hot Encoding Combinations}} \\
\hline
Combinations (Dim) & Cross-domain & In-domain & Average \\
\hline
ABF (204) (Ours)                            & \textbf{6.6863}           & \textbf{4.0}         & \textbf{4.9811}            \\
AB (172)                             & 15.3823&4.5665&9.4701            \\
AF (188)                             & 13.0811&4.0847&7.9097           \\
BF (199)                             & 11.7292&4.1652&7.34            \\
A (156)                              & 20.094&4.7138&11.3221          \\
B (167)                              & 12.7202&4.2781&7.3687           \\
F (183)                              & 11.1229&4.3239&7.2736            \\
Numerical (151)                  & 13.7479&4.0733&8.3435           \\
\hline
\multicolumn{5}{c}{\textbf{Models with Attention Mechanisms}} \\
\hline
Model  & Cross-domain & In-domain & Average & Parameters \\
\hline
ARL (Ours)            & \textbf{6.6863}   & \textbf{4.0}   & \textbf{4.9811}   & 105209 \\
w/o attention         & 8.5095            & 4.0561         & 5.6571            & \textbf{100097} \\
Multi-head Attention  & 9.0952            & 4.0944         & 5.6668            & 163073 \\

\hline
\end{tabular}
\end{adjustbox}
\label{TABLE_COMBINED}
\end{table}

\section{Conclusion}





In this paper, we presented an innovative deep learning model for estimating the energy consumption of 5G BSs, designed to enhance energy-saving efforts in the rapidly evolving 5G networks.


Different from existing methods, our method integrates BSID into the input features to capture different energy fingerprints across various BSs which enables for improved estimation accuracy. This integration is facilitated through an embedding layer, offering a more precise BSID representation with significantly reduced feature dimensions in comparison to One-Hot encoding. Moreover, the introduction of a masked training approach, combined with an ARL attention mechanism, has substantially enhanced the model's generalization capabilities and accuracy.

\section*{Acknowledgment}
The work was supported in part by National Key R\&D Program of China under Grant 2022YFA1003900, in part by Guangdong Major Project of  Basic and Applied Basic Research under Grant 2023B0303000001, in part by Guangdong Young Talent Research Project (Grant No. 2023TQ07A708).

\bibliographystyle{IEEEtran}
\bibliography{ref.bib}

\end{document}